\documentclass[11pt]{article}

\usepackage{mathtools}
\usepackage{amsmath}
\usepackage{amsfonts}
\usepackage{listings}
\usepackage{graphicx}
\usepackage{longtable}
\usepackage[parfill]{parskip}
\usepackage{bbm}
\usepackage{float}
\usepackage{adjustbox}
\usepackage{algorithm}
\usepackage{algorithmic}
\usepackage{caption}
\usepackage{subcaption}
\usepackage{hyperref}

\usepackage{newtxtext, makeidx, multicol, footmisc}

\usepackage{authblk}

\begin{document}

\author[1]{Pierre Miasnikof \thanks{corresponding author: p.miasnikof@mail.utoronto.ca}}
\author[2]{Seo Hong}
\author[1]{Yuri Lawryshyn}

\affil[1]{Dept. of Chemical Engineering and Applied Chemistry, University of Toronto, Toronto, ON, Canada}
\affil[1]{Dept. of Mechanical and Industrial Engineering, University of Toronto, Toronto, ON, Canada}

\title{Graph Clustering Via QUBO and Digital Annealing}

\date{}

\maketitle

\begin{abstract}
This article empirically examines the computational cost of solving a known hard problem, graph clustering, using novel purpose-built computer hardware. We express the graph clustering problem as an intra-cluster distance or dissimilarity minimization problem. We formulate our poblem as a quadratic unconstrained binary optimization problem and employ a novel computer architecture to obtain a numerical solution.  Our starting point is a clustering formulation from the literature. This formulation is then converted to a quadratic unconstrained binary optimization formulation. Finally, we use a novel purpose-built computer architecture to obtain numerical solutions. For benchmarking purposes, we also compare computational performances to those obtained using a commercial solver, Gurobi, running on conventional hardware. Our initial results indicate the purpose-built hardware provides equivalent solutions to the commercial solver, but in a very small fraction of the time required.
\end{abstract}

\section{Introduction}
In this article, we express the graph clustering problem as a quadratic unconstrained binary optimization problem (QUBO), also referred to as an Ising model. Our QUBO model is designed to minimize intra-cluster node-to-node distances or dissimilarities. The overarching goal of this study is to empirically examine the computational cost of solving a known hard problem, graph clustering, by reformulating it as a QUBO problem and solving it numerically on a novel computer architecture, Fujitsu's Digital Annealer (DA). This architecture, the DA, is built specifically for combinatorial optimization problems. 

The work described in this article lies at the intersection of graph clustering, general (metric-space) clustering, combinatorial optimization and high performance computing. While we seek to conduct graph clustering, we reformulate our problem as a metric-space clustering distance minimization problem. Typically, when performing graph clustering, all-pairs vertex-to-vertex distances or dissimilarities are not available. We use a heuristic technique to obtain them. Here, we highlight that for the purpose of this work, vertex-to-vertex distance is meant to mean dissimilarity in vertex neighborhoods, not the typical geodesic distance. Finally, to circumvent the NP-hard nature of the clustering problem, we use a novel computer architecture that is purpose-built to solve QUBO problems.

\begin{figure}
	\centering
	\begin{subfigure}{0.5\textwidth}
		\centering
		\includegraphics[width=\textwidth]{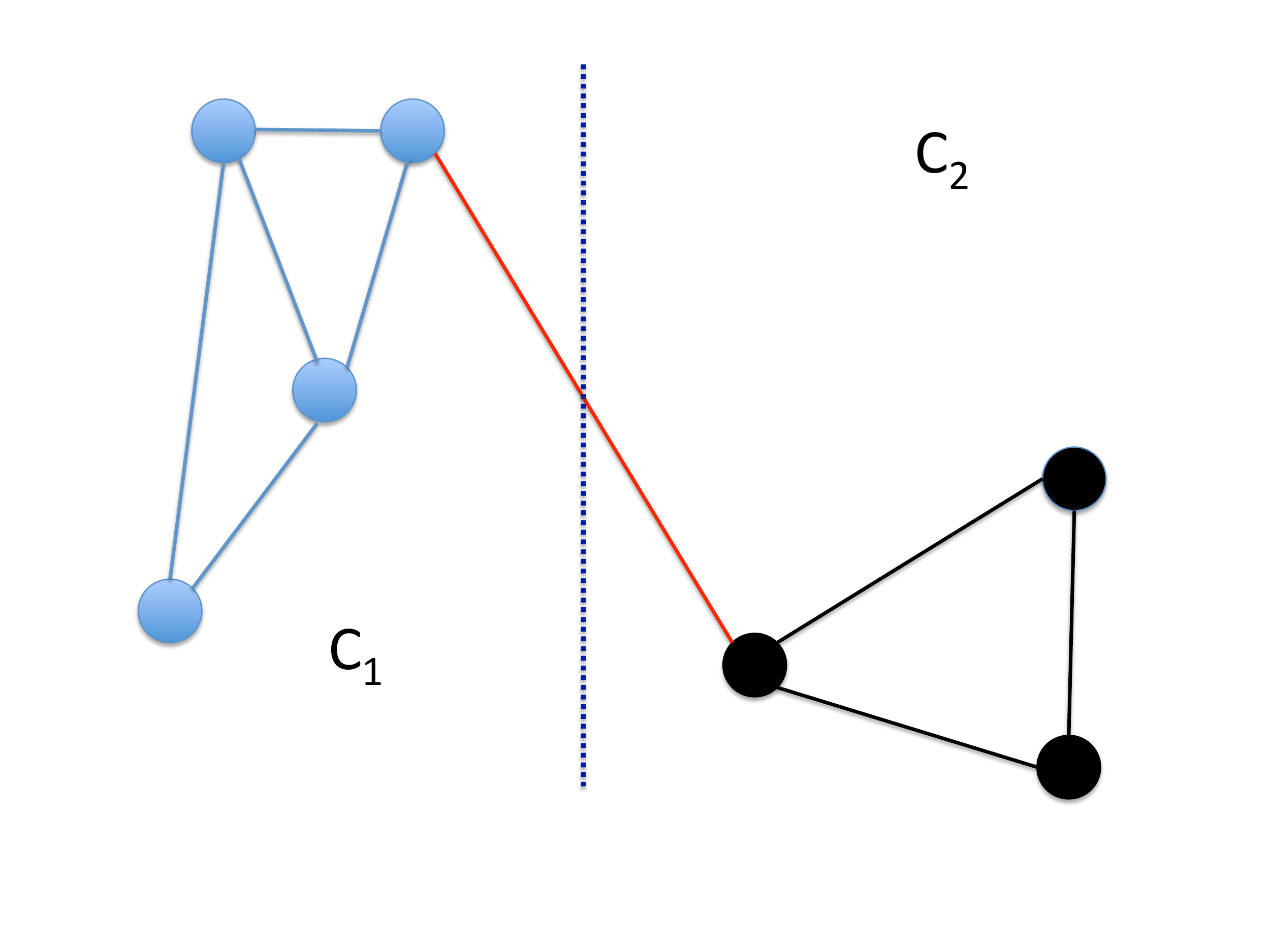}
		\caption{Well Clustered Graph}
	\end{subfigure}%
	~ 
	\begin{subfigure}{0.5\textwidth}
		\centering
		\includegraphics[width=\textwidth]{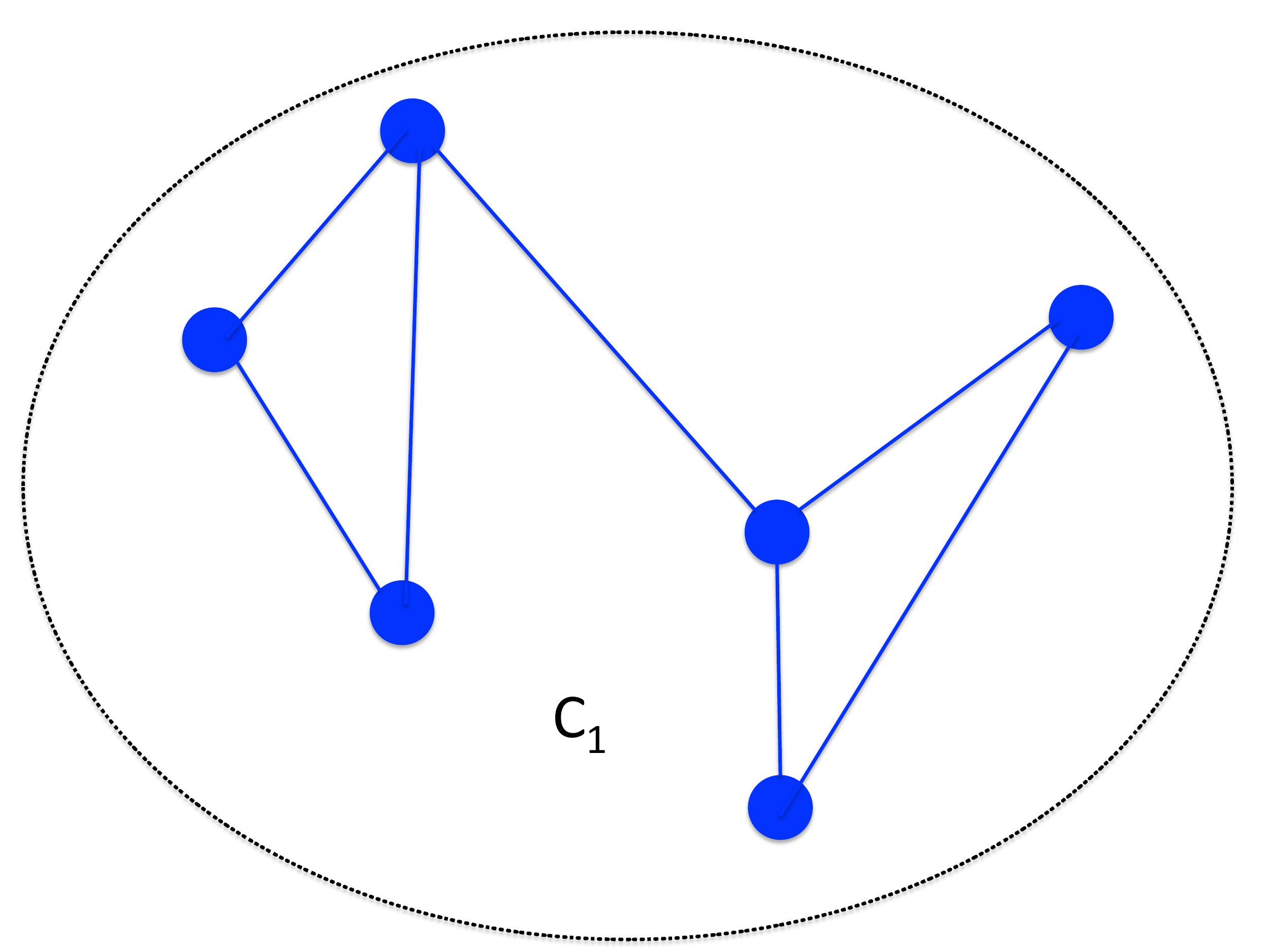}
		\caption{Improperly Clustered Graph}
	\end{subfigure}
	\caption{Examples of Good and Bad Clustering}
	\label{goodNnogood}
\end{figure}

Graph clustering, sometimes called network community detection, is an instance of unsupervised learning. It is a central topic in the field of network science \cite{guideFortunato16} and has even been described as {\it ``one of the most important and challenging problems in network analysis''}, in the very recent literature \cite{LiudaSynthetic2019}. It consists of assigning common labels to vertices considered similar. Typically, similarity is defined by shared connections. Vertices that share more connections are defined as closer, more similar, to each other than to the ones with which they share fewer connections. Successful clustering results in vertices grouped into densely connected induced subgraphs (e.g., \cite{PMEtAlWAW18,PMEtAlWAW19,PMEtAlCplxNets20}). Figure~\ref{goodNnogood} shows an example of a successful and an unsuccessful clustering. Unfortunately, graph clustering is also an NP-hard problem \cite{Schaeffer2007,FortunatoLong2010}.

A formal definition of graph clusters (or network communities) remains a matter of debate in the literature (e.g., \cite{Schaeffer2007,guideFortunato16,LiudaSynthetic2019}) and a topic beyond the scope of this article. However, most authors agree that a cluster can be described as a dense subgraph within a sparser graph (e.g., \cite{FortunatoLong2010,YangLesko2012,modWAW2016,EuroComb2017}). Here, it must also be mentioned that clusters are not necessarily cliques. In fact, in most cases, clusters are not cliques \cite{guideFortunato16}.

It is also important to highlight that not all graphs are clusterable. In many cases, clusters do not offer a meaningful summary description of a graph's structure. For example, clusters are arguably uninformative in the case of complete graphs. In fact, the topic of clusterability, the assessment of a graph's suitability to a clustering exercise, has received some attention in the recent literature \cite{GaoLaff2017Stat,GaoLaff2017Prob,ChipEtAl2018,PMEtAlLION19}. For the purpose of this article, we restrict our attention to clusterable, undirected, unweighted and weighted graphs, with no self loops or multiple edges. 

While there are various approaches to graph clustering, we focus on an optimization-based problem formulation. Our formulation consists of assigning cluster labels by minimizing total intra-cluster distance or dissimilarity between nodes within each cluster. As mentioned earlier, vertex-to-vertex distances or (dis)similarity measures are typically not available. We use a heuristic described in Fortunato's exhaustive work to obtain these distances \cite{FortunatoLong2010}. These distances represent measures of vertex neighnorhood dissimilarity, dissimilarity in connectivity patterns. Here, our graph clustering problem is transformed into a more common (metric) clustering technique, intra-cluster distance or disssimilarity minimization. 

Most authors who have used optimization-based approaches maximize modularity. Our approach is more flexible and allows us to circumvent modularity's many shortcomings. These shortcomings have been well-documented in the literature (e.g., \cite{ResolLimitFortunato2007,AckerBD08,PMEtAlWAW18,PMEtAlWAW19,PMEtAlCplxNets20}). Furthermore, we choose to formulate our problem as a QUBO problem, in order to overcome computational intractability and benefit from new hardware developments. Proceeding in this way allows us to take advantage of new purpose-built computational hardware, such as  the DA, to obtain numerical solutions.

The remainder of this article is organized as follows. In Section~\ref{litReview}, we briefly examine the previous work done in both the fields of graph clustering and QUBO problems. In Section~\ref{methods}, we describe our formulation of the graph clustering problem as a distance minimization problem and the test graphs we use to illustrate our technique and compare computation performances. A brief description of the DA is provided in Section~\ref{DA}. Numerical results are presented in Section~\ref{results}.

\section{Previous Work} \label{litReview}
Although it falls under the broad umbrella of clustering and unsupervised learning \cite{ESL09}, graph clustering is a distinct field. The main distinction lies in the fact that graphs are not typically in metric space. All-pairs distances are not typically available. This difference distinguishes graph clustering from the more traditional clustering problems which are set in metric-space, such as K-means, for example \cite{ESL09}.

A thorough review of the graph clustering literature is beyond the scope of this article. For a very broad and thorough overview of the field, we refer the reader to the foundational work of Schaeffer \cite{Schaeffer2007}, Fortunato \cite{FortunatoLong2010} and the recent contribution by Fortunato and Hric \cite{guideFortunato16}. Nevertheless, we note the existence of various competing graph clustering techniques, built on very different mathematical foundations. The main competing approaches to graph clustering are
\begin{itemize}
\item Spectral (e.g., \cite{Lux})
\item Markov (e.g., \cite{MCMain})
\item Optimization 
\begin{itemize}
\item Modularity maximization (e.g., \cite{AloiseEtAl2012,GRASP_Pitsoulis,guideFortunato16})
\item Other objective functions (e.g., \cite{FanPard2010CND,FanPard2010LQ,LanciEtAlStat2011,FanZhengPardalosIntervals2012}).
\end{itemize}
\end{itemize}
In spite of its known shortcomings (e.g., \cite{ResolLimitFortunato2007,AckerBD08,PMEtAlWAW18,PMEtAlWAW19,PMEtAlCplxNets20}), the most common optimization formulation is modularity maximization. However, in contrast,  Fan and Pardalos maximize intra-cluster vertex similarity (minimize dissimilarity) \cite{FanPard2010CND,FanPard2010LQ}. The work in this article closely follows these authors' framework.

Of course, it is important to also compare optimization-based approaches to other commonly used graph clustering techniques. Here, we must point out that spectral methods come with a heavy computational cost and do not work well on larger instances. This scale limitation was noted by Schaeffer \cite{Schaeffer2007}. Although some authors' more recent algorithms are described as ``faster and more accurate'', they still carry a heavy computational cost (e.g., \cite{FastScore2015}). More importantly, spectral methods have been described as ill-suited to sparse graphs \cite{guideFortunato16}. Unfortunately, clusterable graphs, graphs whose structure can be meaningfully described using clusters, are typically sparse. 

Markov-based techniques revolve around simulations of random walks over the graph. Such simulations require numerous matrix multiplications. Additionally, Markov clustering also requires various element-wise and row operations. An appealing feature of Markov clustering is that it doesn't require the number of clusters as a parameter input. While this feature may be advantageous in cases where a reasonable guess for the number of clusters is not known, in most cases domain knowledge does provide clues about this number. Since it is known that algorithms that do not require the number of clusters as an input parameter have been found to be less accurate than those that do require it \cite{guideFortunato16}, this initially appealing feature of Markov clustering may be a weakness in most cases.

On the other hand, optimization-based approaches lend themselves very well to approximate solution techniques, which carry a lower computational cost. Indeed, because of the NP-hardness of the graph clustering problem \cite{Schaeffer2007,FortunatoLong2010}, solving these and other types of combinatorial optimization problems is often successfully done via (meta-)heuristic solution techniques (e.g., \cite{Papadimitriou98}), which explore subsets of the solution space. In the specific case of graph clustering, many authors have made use of meta-heuristic optimization techniques (e.g., \cite{AloiseEtAl2012,GRASP_Pitsoulis,GenClust14}), in order to find approximate solutions and overcome the NP-hard nature of the problem. Additionally, meta-heuristic optimization techniques are easily parallelizable and well suited to implementation on high performance computing platforms.  

Numerous NP-hard problems have also been reformulated as Ising (QUBO) problems \cite{Fu86,IsingForm2014}. Such reformulations allow the implementation on massively parallel purpose-built hardware which yield solutions using simulated annealing \cite{accel2017,applic2019,physInspired2019,Glover2019}. In fact, the graph partitioning problem is one of the original problems at the intersection of Ising modeling and the study of NP-complete problems \cite{IsingForm2014}.

\section{Methods} \label{methods}
We formulate an optimization problem that assigns cluster labels to nodes in a manner that minimizes intra-cluster dissimilarity. We define dissimilarity as the sum of intra-cluster node-to-node distances. Our formulation is a variation of the work of Fan and Pardalos \cite{FanPard2010CND,FanPard2010LQ}. 

The distinguishing feature of our approach comes from our choice of dissimilarity measure, our handling of constraints and our computational solution technique. We use a heuristic technique described by Fortunato \cite{FortunatoLong2010} to obtain vertex-to-vertex distances from the graph's adjacency matrix. We then formulate our problem as a QUBO problem and use the DA to solve it \cite{applic2019,physInspired2019}. We also compare our computational results to those obtained using the Gurobi commercial solver (\url{http://www.gurobi.com}).

\subsection{Objective Function}
Here, we describe our model's objective function. As mentioned earlier, the goal of this formulation is to minimize intra-cluster dissimilarity. This intra-cluster dissimilarity is defined as the sum of intra-cluster node-to-node distances.

Our objective function, which we minimize, is the sum of intra-cluster dissimilarities over all clusters. For a graph with $\vert V \vert = N$ vertices to be partitioned into a set of clusters $C$ (where, $\vert C \vert = K$) , it is defined as:
\begin{eqnarray}
f_o &=& \sum_{i=1}^N \sum_{j=i+1}^N \sum_{k=1}^{K} d_{ij} x_{ik} x_{jk} \; .
\end{eqnarray}
The binary decision variables $x_{ik},x_{jk} \in \{0,1\}$ are defined as:
\begin{eqnarray}
x_{ik} &=& \left\{ \begin{array}{l l}
1 & \text{if vertex $i$ is assigned to cluster $k$} \\
0 & \text{otherwise} \; .
\end{array} \right. \label{binCon}
\end{eqnarray}
The coefficients $d_{ij} (\ge 0)$ denote the distance (dissimilarity) between vertices $i$ and $j$. Also, in our formulation, the number of clusters into which we partition the graph, $K$, is an input parameter. Requiring the number of clusters as an input parameter is consistent with many common clustering techniques, such as K-means, for example \cite{ESL09}. As mentioned earlier, proceeding this way has also been found to produce better clustering results than determining the number of clusters through a clustering algorithm that doesn't require it as a parameter input \cite{guideFortunato16}. 

\subsection{Constraints} \label{constrs}
In addition to the binary constraints on the decision variables defined in Equation~(\ref{binCon}), we formulate one additional set of constraints and an $L_2$ regularizer. The set of constraints in Equation~(\ref{oneClust}), one for each vertex, ensures all vertices are assigned to exactly one cluster. The regularizer, in Equation~(\ref{L2}), ensures clusters are similarly sized. In the expressions below, the input parameter $\bar{U}$ specifies the desired approximate cluster size, $V$ denotes the set of vertices, $C$ the set of clusters and $K = \vert C \vert$ its cardinality:
\begin{eqnarray}
h_i &=& \sum_{k=1}^K x_{ik} = 1 \quad \forall i \in V \label{oneClust} \\
R_k &=& \left( \left( \sum_{i=1}^N x_{ik} \right) - \bar{U} \right)^2 \quad \forall k \in C \label{L2} \, .
\end{eqnarray}

\subsection{Inter-vertex Distances $(d_{ij})$}
Graphs are typically not in metric space, inter-vertex distances are not trivially defined. We use the vertex-to-vertex distance definition presented by Fortunato \cite{FortunatoLong2010} but attributed to Burt \cite{BurtDist76} and to  Wasserman and Faust \cite{wasserman_faust_1994}:
\begin{eqnarray}
d_{ij} &=& \sqrt{ \sum_{\ell \ne i,j} \left( A_{i\ell} - A_{j\ell} \right)^2 } \; . \label{dists}
\end{eqnarray}

$A_{i\ell}$ is the element at the intersection of the $i\text{-}th$ row and $\ell\text{-}th$ column in the graph's adjacency matrix $(A)$. 

As mentioned earlier, our goal is to label (cluster) vertices so they form dense induced subgraphs. Therefore, the distance which we minimize is a measure of vertex-to-vertex pairwise neighborhood dissimilarity, not geodesic or Euclidian distance. Vertices that share fewer connections are consider more distant to each other, than vertices that share a larger number of neighbors. We choose this technique to obtain all-pairs distances, because of its simplicity and for the purpose of illustration. However, there are a variety of different vertex-to-vertex distance alternatives, like Jaccard distances, for example \cite{JaccOrig}. It must also be noted that our formulation is independent of the chosen distance measure.

\subsection{QUBO Formulation}
A QUBO formulation is, by definition, unconstrained. The only constraints are the inherent binary constraints imposed on the decision variables. Constrained optimization problems, such as ours, are expressed in QUBO form by including quadratic penalties in the objective function \cite{Glover2019}.

We formulate two different optimization models. The first model, model 1,  is shown in Equation~(\ref{prob1}). It only has two sets of constraints, the binary constraints on the decision variables and the cluster membership constraints that ensure each vertex belongs to exactly one cluster. The latter are the sum of the squared  $(h_i -1)$ shown in Equation~\ref{oneClust}, multiplied by a positive penalty coefficient $P$.

The second model, model 2, is shown in Equation~(\ref{prob2}). It has all the constraints of the first model and also includes an $L_2$ regularizer. The regularizer is the sum of the $R_k$ terms in Equation~\ref{L2}, multiplied by a positive penalty coefficient $\lambda$. It pushes cluster sizes towards uniformity, towards a size of approximately $\bar{U}$.

The two QUBO models, model 1 and model 2, are given by:
\begin{align}
 \underset{x_{ik},x_{jk}}{\text{minimize}} & \left\{ \sum_{i=1}^N \sum_{j=i+1}^N \sum_{k=1}^{K} d_{ij} x_{ik} x_{jk}  + \sum_{i=1}^N P \left( \left( \sum_{k=1}^K x_{ik} \right) - 1 \right)^2 \right\}  \label{prob1} \\
 \underset{x_{ik},x_{jk}}{\text{minimize}} & \left\{ \sum_{i=1}^N \sum_{j=i+1}^N \sum_{k=1}^{K} d_{ij} x_{ik} x_{jk}  + \sum_{i=1}^N P \left( \left( \sum_{k=1}^K x_{ik} \right) - 1 \right)^2 \right. \nonumber  \\
&\left. + \sum_{k=1}^K \lambda  \left( \left( \sum_{i=1}^N x_{ik} \right) - \bar{U} \right)^2  \right\} \label{prob2} \\
& \left( x_{ik} \in \{0,1\},  \; \forall i \in V, \forall k \in C \right) \, . \nonumber
\end{align}
The parameters $P$ and $\lambda$ specify the penalty weights for the cluster membership constraints and $L_2$ regularizers, respectively. As described earlier, the binary decision variables are denoted as $x_{ik}$, the distances separating nodes is denoted by $d_{ij}$, the set of vertices is $V$ and the set of clusters is $C$.

According to Glover et al. \cite{Glover2019}, penalty coefficients for constraints should be in the interval $[0.75,1.5]$. However, our numerical experiments revealed that even a penalty coefficient of 1.5 was insufficient. We found that a penalty coefficient of at least $P= 16$ was necessary to ensure the constraint was not violated. On the other hand, we set $\lambda = 0.75$, because we just want a soft constraint on cluster cardinalities. We want to ensure nodes are always strictly assigned to exactly one cluster, while we only want to have clusters of roughly equal sizes. For the purpose of this experiment, we arbitrarily set $\bar{U} = \frac{N}{K}$, so that all clusters would contain roughly $\frac{1}{K}$ of all nodes. This parameter can be adjusted to suit domain knowledge and can be specified to accommodate clusters of varying sizes.

\subsection{Preventing Common Degeneracies}
Graph clustering algorithms may suffer from two commonly observed degeneracies: mega-clusters and micro-clusters. The first degeneracy refers to instances where clustering algorithms lump the bulk or the totality of vertices into one or just a few clusters. This phenomenon is especially common in the case of modularity maximization. It is a consequence of a well-known weakness of modularity \cite{ResolLimitFortunato2007,FortunatoLong2010,Laar14}. The second is the tendency of algorithms to identify clusters consisting of just a small number of vertices, where clusters should reasonably be expected to contain more vertices. This situation is often characterized by a large number of very small, very strongly clustered clusters, often accompanied by a small number of large poorly clustered ``garbage collector'' clusters where the remaining vertices are lumped together. Our second model prevents these common degeneracies, with the addition of the $L_2$ regularizer.

\subsection{Test Data \& Test Cases} \label{testData}
We implement our models on five different synthetic graphs. Each graph's characteristics are described in Table~\ref{graphChars}, below. Our graphs are generated using the stochastic block model \cite{SBMOrig83}, as implemented in the NetworkX simulation function \cite{Networkx}. We generate smaller graphs containing eight clusters, but fewer vertices overall and  larger graphs containing only four clusters but a higher total number of vertices. The resulting number of vertices is shown in the second column.

Our graphs also have varying levels of noise in their connectivity patterns. Noise is introduced in the form of a broader range of intra-cluster and inter-cluster edge probabilities. For example, the Low-Noise-Large (L4) graph was generated using intra-cluster edge probabilities drawn randomly from a uniform distribution on the interval $(0.9,1)$ and inter-cluster edge probabilities drawn from a uniform distribution on the interval $(0,0.2)$. Meanwhile, the Very-High-Noise-Small (VH8) graph was generated using intra-cluster edge probabilities drawn randomly from a uniform distribution on the interval $(0.7,1)$ and inter-cluster edge probabilities drawn from a uniform distribution on the interval $(0,0.55)$.
\begin{table}[]
\centering
\caption{Graph Characteristics} \label{graphChars}
\begin{tabular}{l c c c l}
\hline
{\bf Name}  & {\bf Verts$(\vert V \vert)$} & {\bf Clusts$(K)$} & {\bf Intra Pr} & {\bf Inter Pr}  \\
\hline
Low-Noise-Large (L4)   & 247   & 4      & U(0.9,1)   & U(0,0.2)    \\
Low-Noise-Small (L8)   & 120   & 8      & U(0.9,1)   & U(0,0.2)   \\
High-Noise-Large (H4) & 253   & 4      & U(0.7,1)   & U(0,0.4)   \\
High-Noise-Small (H8) & 127   & 8      & U(0.7,1)   & U(0,0.4)   \\
Very-High-Noise-Small (VH8) & 122 & 8 & U(0.7,1) & U(0,0.55) \\
\hline             
\end{tabular}
\end{table}

For each of these graphs, we minimize the functions shown in Equations~\ref{prob1} and~\ref{prob2} using both the DA and a commercial solver. This solver was run on a XEON based system with two 18 core E5-2697 v4 processors and 256 GB of 2133 Mhz ECC RAM, running MS Windows Server 2012R2. Numerical results are shown in Section~\ref{results}.

\subsection{Fujitsu Digital Annealer} \label{DA}
The Fujitsu Digital Annealer (DA) is a purpose-built computer architecture designed to efficiently solve combinatorial optimization problems \cite{applic2019,website}. Specifically, it is designed to solve fully connected Ising model problems \cite{accel2017} of up to 1,024 bits. (A more recent version allows for problem instances of up to 8,192 bits.)

As mentioned earlier, many NP-hard problems can be expressed using the Ising model \cite{IsingForm2014,physInspired2019}. While adiabatic quantum optimization (AQO) is unlikely to yield polynomial-time solutions to all NP-complete problems, it is likely to be superior to classical algorithms solved on traditional computing platforms \cite{IsingForm2014}. In fact, such performance improvements have already been documented \cite{accel2017}. The DA exploits the hypothesis of AQO superiority by emulating qubits on a digital chip \cite{physInspired2019} (\url{https://www.fujitsu.com/global/digitalannealer/superiority/}).

\section{Results} \label{results}

Tables~\ref{res1} and~\ref{res2} display our numerical results. The columns contain:
\begin{enumerate}
\item Graph names
\item Run times for the DA (in seconds)
\item Run times required for the commercial solver to reach same objective function value (in seconds)
\item Ratio of DA run times over commercial solver run times to reach same objective function value
\item Best objective function value returned by DA after 3.5 seconds
\item Best objective function value returned by the commercial solver after 360 seconds
\item Ratio of DA best over commercial solver best .
\end{enumerate}

\begin{table}[]
\tiny
\centering
\caption{Run Times and Objective Function Values, Model 1} \label{res1}
\begin{tabular}{| l | c | c | c | c | c | c |}
\hline
{\bf Name}  & {\bf DA Time (s)} & {\bf Gi Time (s)} & {\bf Time DA/Time Gi} & {\bf Best DA (3.5s)} & {\bf Best Gi (360s)} & {\bf Obj DA/Obj Gi Best} \\
\hline
Low-Noise-Large (L4)   & 3.5   & 360 & 0.01 & 44803 & 44803 & 1.00 \\
Low-Noise-Small (L8)   & 3.5  & 97 & 0.04 & 67594 & 67594 & 1.00 \\
High-Noise-Large (H4)  & 3.5  & 41 & 0.09 & 3690 & 3690 & 1.00 \\
High-Noise-Small (H8)  & 3.5  & 178 & 0.02 & 5740 & 5713 & 1.00 \\
Very-High-Noise-Small (VH8) & 3.6  & 134 & 0.03 & 5761 & 5755 & 1.00 \\
\hline
\end{tabular}
\end{table}

\begin{table}[]
\tiny
\centering
\caption{Run Times and Objective Function Values, Model 2} \label{res2}
\begin{tabular}{| l | c | c | c | c | c | c |}
\hline
{\bf Name}  & {\bf DA Time (s)} & {\bf Gi Time (s)} & {\bf Time DA/Time Gi} & {\bf Best DA (3.5s)} & {\bf Best Gi (360s)} & {\bf Obj DA/Obj Gi Best} \\
\hline
Low-Noise-Large (L4)   & 3.5  & 49 & 0.07 & 44943 & 44941 & 1.00 \\
Low-Noise-Small (L8)   & 3.5  & 49 & 0.07 & 67649 & 67649 & 1.00\\
High-Noise-Large (H4)  & 3.5  & 178 & 0.02 & 3697 & 3697  & 1.00\\
High-Noise-Small (H8)  & 3.5  & 127 & 0.03 & 5771 & 5766  & 1.00\\
Very-High-Noise-Small (VH8) & 3.5  & 102 & 0.03 & 5767 & 5748 & 1.00\\
\hline
\end{tabular}
\end{table}

Our results show the DA offers an equivalent solution but with computation times that are orders of magnitude smaller. In the last column of both tables, we see the objective function values returned by the DA are roughly equal to those returned by the commercial solver. In the fourth column, we see the DA reached equivalent results but with only a small fraction of the time it took the commercial solver.

\section{Conclusion}
In this article, we formulate the graph clustering problem as an intra-cluster distance minimization problem, using the Ising model. We then proceed to solve our problem using the DA. Our initial results show dramatically shorter run times than solving the same problems using a commercial solver.

Although commercial solver computation times were relatively short, they were orders of magnitude greater than DA computation times. However, our comparisons were limited by current hardware architecture. We reasonably expect the DA's computational advantage to remain in cases of much larger graphs. This advantage may prove to be critical in cases where graph sizes render commercial solver run times unreasonably long and possibly too long for uses in real-world situations. 

Future work will focus on examining the effect of different vertex-to-vertex distance measures and the effect of noise in connectivity patterns on solution times and solution quality. With the recent introduction of a second-generation DA, we will also examine larger graphs.

\section*{Acknowledgements}
The authors would like to thank Fujitsu Laboratories Ltd and Fujitsu Consulting (Canada) Inc for providing financial support and access to the Digital Annealer at the University of Toronto.

\bibliographystyle{spmpsci}  

\bibliography{consolidatedBib}

\begin{thebibliography}{10}
\providecommand{\url}[1]{{#1}}
\providecommand{\urlprefix}{URL }
\expandafter\ifx\csname urlstyle\endcsname\relax
  \providecommand{\doi}[1]{DOI~\discretionary{}{}{}#1}\else
  \providecommand{\doi}{DOI~\discretionary{}{}{}\begingroup
  \urlstyle{rm}\Url}\fi

\bibitem{AckerBD08}
Ackerman, M., Ben-David, S.: Measures of clustering quality: A working set of
  axioms for clustering.
\newblock Advances in Neural Information Processing Systems 21 - Proceedings of
  the 2008 Conference pp. 121--128 (2008)

\bibitem{AloiseEtAl2012}
Aloise, D., Caporossi, G., Hansen, P., Liberti, L., Perron, S., Ruiz, M.:
  Modularity maximization in networks by variable neighborhood search.
\newblock In: D.A. Bader, H.~Meyerhenke, P.~Sanders, D.~Wagner (eds.) Graph
  Partitioning and Graph Clustering, 10th {DIMACS} Implementation Challenge
  Workshop, Georgia Institute of Technology, Atlanta, GA, USA, February 13-14,
  2012. Proceedings, pp. 113--128 (2012).
\newblock \urlprefix\url{http://www.ams.org/books/conm/588/11705}

\bibitem{physInspired2019}
{Aramon}, M., {Rosenberg}, G., {Valiante}, E., {Miyazawa}, T., {Tamura}, H.,
  {Katzgraber}, H.: {Physics-Inspired Optimization for Quadratic Unconstrained
  Problems Using a Digital Annealer}.
\newblock Frontiers in Physics \textbf{7}, 48 (2019).
\newblock \doi{10.3389/fphy.2019.00048}.
\newblock
  \urlprefix\url{https://www.frontiersin.org/article/10.3389/fphy.2019.00048}

\bibitem{BurtDist76}
Burt, R.: {Positions in Networks*}.
\newblock Social Forces \textbf{55}(1), 93--122 (1976)

\bibitem{ChipEtAl2018}
{Chiplunkar}, A., {Kapralov}, M., {Khanna}, S., {Mousavifar}, A., {Peres}, Y.:
  {Testing Graph Clusterability: Algorithms and Lower Bounds}.
\newblock ArXiv e-prints  (2018)

\bibitem{MCMain}
van Dongen, S.: Graph clustering by flow simulation.
\newblock Ph.D. thesis, Faculteit Wiskunde en Informatica, Universiteit Utrecht
  (2000)

\bibitem{FanPard2010LQ}
Fan, N., Pardalos, P.M.: Linear and quadratic programming approaches for the
  general graph partitioning problem.
\newblock J. of Global Optimization \textbf{48}(1), 57--71 (2010).
\newblock \doi{10.1007/s10898-009-9520-1}.
\newblock \urlprefix\url{http://dx.doi.org/10.1007/s10898-009-9520-1}

\bibitem{FanPard2010CND}
Fan, N., Pardalos, P.M.: Robust optimization of graph partitioning and critical
  node detection in analyzing networks.
\newblock In: Proceedings of the 4th International Conference on Combinatorial
  Optimization and Applications - Volume Part I, COCOA'10, pp. 170--183.
  Springer-Verlag, Berlin, Heidelberg (2010).
\newblock \urlprefix\url{http://dl.acm.org/citation.cfm?id=1940390.1940405}

\bibitem{FanZhengPardalosIntervals2012}
Fan, N., Zheng, Q.P., Pardalos, P.M.: Robust optimization of graph partitioning
  involving interval uncertainty.
\newblock Theoretical Computer Science - TCS \textbf{447}, 53--61 (2012).
\newblock \doi{10.1016/j.tcs.2011.10.015}

\bibitem{FortunatoLong2010}
{Fortunato}, S.: {Community detection in graphs}.
\newblock Physics Reports \textbf{486}, 75--174 (2010).
\newblock \doi{10.1016/j.physrep.2009.11.002}.
\newblock \urlprefix\url{http://adsabs.harvard.edu/abs/2010PhR...486...75F}

\bibitem{ResolLimitFortunato2007}
Fortunato, S., Barth\'{e}lemy, M.: Resolution limit in community detection.
\newblock Proceedings of the National Academy of Sciences \textbf{104}(1),
  36--41 (2007).
\newblock \doi{10.1073/pnas.0605965104}.
\newblock \urlprefix\url{http://www.pnas.org/content/104/1/36.abstract}

\bibitem{guideFortunato16}
Fortunato, S., Hric, D.: Community detection in networks: A user guide.
\newblock Physics Reports \textbf{659}, 1--44 (2016).
\newblock \doi{10.1016/j.physrep.2016.09.002}

\bibitem{Fu86}
Fu, Y., Anderson, P.W.: Application of statistical mechanics to {NP}-complete
  problems in combinatorial optimisation.
\newblock Journal of Physics A: Mathematical and General \textbf{19}(9),
  1605--1620 (1986)

\bibitem{website}
Fujitsu:  \urlprefix\url{https://www.fujitsu.com/global/digitalannealer/}

\bibitem{GaoLaff2017Stat}
{Gao}, C., {Lafferty}, J.: {Testing for Global Network Structure Using Small
  Subgraph Statistics}.
\newblock ArXiv e-prints  (2017)

\bibitem{GaoLaff2017Prob}
{Gao}, C., {Lafferty}, J.: {Testing Network Structure Using Relations Between
  Small Subgraph Probabilities}.
\newblock ArXiv e-prints  (2017)

\bibitem{Glover2019}
{Glover}, F., {Kochenberger}, G., {Du}, Y.: {A Tutorial on Formulating and
  Using QUBO Models}.
\newblock arXiv e-prints arXiv:1811.11538 (2018)

\bibitem{Networkx}
Hagberg, A., Schult, D., Swart, P.: {Exploring Network Structure, Dynamics, and
  Function using NetworkX}.
\newblock In: G.~Varoquaux, T.~Vaught, J.~Millman (eds.) Proceedings of the 7th
  Python in Science Conference, pp. 11 -- 15. Pasadena, CA USA (2008)

\bibitem{ESL09}
Hastie, T., Tibshirani, R., Friedman, J.: {The Elements of Statistical
  Learning, Second Edition: Data Mining, Inference, and Prediction}, 2nd ed.
  2009. edn.
\newblock Springer Series in Statistics. Springer (2009)

\bibitem{SBMOrig83}
Holland, P.W., Laskey, K.B., Leinhardt, S.: Stochastic blockmodels: First
  steps.
\newblock Social Networks \textbf{5}(2), 109 -- 137 (1983).
\newblock \doi{https://doi.org/10.1016/0378-8733(83)90021-7}.
\newblock
  \urlprefix\url{http://www.sciencedirect.com/science/article/pii/0378873383900217}

\bibitem{JaccOrig}
Jaccard, P.: \'{E}tude de la distribution florale dans une portion des {A}lpes
  et du {J}ura.
\newblock Bulletin de la Soci\'{e}t\'{e} Vaudoise des Sciences Naturelles
  \textbf{37}, 547--579 (1901).
\newblock \doi{10.5169/seals-266450}

\bibitem{FastScore2015}
Jin, J.: {FAST COMMUNITY DETECTION BY SCORE}.
\newblock The Annals of Statistics \textbf{43} (2015)

\bibitem{GenClust14}
Kazakovtsev, L., Antamoshkin, A.: Genetic algorithm with fast greedy heuristic
  for clustering and location problems.
\newblock Informatica (Slovenia) \textbf{38}(3) (2014).
\newblock
  \urlprefix\url{http://www.informatica.si/index.php/informatica/article/view/704}

\bibitem{Laar14}
Laarhoven, T.V., Marchiori, E.: Axioms for graph clustering quality functions.
\newblock J. Mach. Learn. Res. \textbf{15}(1), 193--215 (2014).
\newblock \urlprefix\url{http://dl.acm.org/citation.cfm?id=2627435.2627441}

\bibitem{LanciEtAlStat2011}
{Lancichinetti}, A., {Radicchi}, F., {Ramasco}, J.J., {Fortunato}, S.: {Finding
  Statistically Significant Communities in Networks}.
\newblock PLoS ONE \textbf{6}, e18961 (2011).
\newblock \doi{10.1371/journal.pone.0018961}

\bibitem{IsingForm2014}
{Lucas}, A.: {Ising formulations of many NP problems}.
\newblock Frontiers in Physics \textbf{2}, 5 (2014).
\newblock \doi{10.3389/fphy.2014.00005}

\bibitem{Lux}
von Luxburg, U.: {A Tutorial on Spectral Clustering}.
\newblock CoRR \textbf{abs/0711.0189} (2007).
\newblock \urlprefix\url{http://arxiv.org/abs/0711.0189}

\bibitem{PMEtAlLION19}
Miasnikof, P., Prokhorenkova, L., Shestopaloff, A., Raigorodskii, A.: {A
  Statistical Test of Heterogeneous Subgraph Densities To Assess
  Clusterability}.
\newblock Accepted, to appear in Springer Lecture Notes in Computer Science
  (2019)

\bibitem{PMEtAlWAW18}
Miasnikof, P., Shestopaloff, A., Bonner, A., Lawryshyn, Y.: A Statistical
  Performance Analysis of Graph Clustering Algorithms, chap.~11.
\newblock Lecture Notes in Computer Science. Springer Nature (2018)

\bibitem{PMEtAlWAW19}
{Miasnikof}, P., {Shestopaloff}, A., {Bonner}, A., {Lawryshyn}, Y., {Pardalos},
  P.: {A Statistical Density-Based Analysis of Graph Clustering Algorithm
  Performance}.
\newblock arXiv e-prints arXiv:1906.02366 (2019)

\bibitem{PMEtAlCplxNets20}
{Miasnikof}, P., {Shestopaloff}, A., {Bonner}, A., {Lawryshyn}, Y., {Pardalos},
  P.: {A Density-Based Statistical Analysis of Graph Clustering Algorithm
  Performance}.
\newblock Journal of Complex Networks, to appear  (2020)

\bibitem{GRASP_Pitsoulis}
Nascimento, M., Pitsoulis, L.: {Community detection by modularity maximization
  using GRASP with path relinking}.
\newblock Computers and Operations Research \textbf{40}, 3121--3131 (2013)

\bibitem{Papadimitriou98}
Papadimitriou, C., Steiglitz, K.: Combinatorial Optimization: Algorithms and
  Complexity.
\newblock Dover Books on Computer Science. Dover Publications (1998).
\newblock \urlprefix\url{https://books.google.ca/books?id=u1RmDoJqkF4C}

\bibitem{LiudaSynthetic2019}
{Prokhorenkova}, L.: {Using synthetic networks for parameter tuning in
  community detection}.
\newblock arXiv e-prints arXiv:1906.04555 (2019)

\bibitem{modWAW2016}
Prokhorenkova, L.O., Pra{\l}at, P., Raigorodskii, A.: Modularity of complex
  networks models.
\newblock In: A.~Bonato, F.~Graham., P.~Pra{\l}at (eds.) Algorithms and Models
  for the Web Graph, pp. 115--126. Springer International Publishing, Cham
  (2016)

\bibitem{EuroComb2017}
Prokhorenkova, L.O., Pra{\l}at, P., Raigorodskii, A.: Modularity in several
  random graph models.
\newblock Electronic Notes in Discrete Mathematics \textbf{61}, 947--953
  (2017).
\newblock \doi{https://doi.org/10.1016/j.endm.2017.07.058}.
\newblock
  \urlprefix\url{http://www.sciencedirect.com/science/article/pii/S1571065317302238}.
\newblock The European Conference on Combinatorics, Graph Theory and
  Applications (EUROCOMB'17)

\bibitem{applic2019}
{Sao}, M., {Watanabe}, H., {Musha}, Y., {Utsunomiya}, A.: {Application of
  Digital Annealer for Faster Combinatorial Optimization}.
\newblock Fujitsu Scientific and Technical Journal \textbf{55}(2), 45--51
  (2019)

\bibitem{Schaeffer2007}
Schaeffer, S.E.: Survey: Graph clustering.
\newblock Comput. Sci. Rev. \textbf{1}(1), 27--64 (2007).
\newblock \doi{10.1016/j.cosrev.2007.05.001}.
\newblock \urlprefix\url{http://dx.doi.org/10.1016/j.cosrev.2007.05.001}

\bibitem{accel2017}
{Tsukamoto}, S., {Takatsu}, M., {Matsubara}, S., {Tamura}, H.: {An Accelerator
  Architecture for Combinatorial Optimization Problems}.
\newblock Fujitsu Scientific and Technical Journal pp. 8--13 (2017)

\bibitem{wasserman_faust_1994}
Wasserman, S., Faust, K.: Social Network Analysis: Methods and Applications.
\newblock Structural Analysis in the Social Sciences. Cambridge University
  Press (1994).
\newblock \doi{10.1017/CBO9780511815478}

\bibitem{YangLesko2012}
Yang, J., Leskovec, J.: {Defining and Evaluating Network Communities based on
  Ground-truth}.
\newblock CoRR \textbf{abs/1205.6233} (2012).
\newblock \urlprefix\url{http://arxiv.org/abs/1205.6233}

\end{thebibliography}

\end{document}